\documentclass[12pt]{article}

\usepackage{epsf}

\newcommand{\bea}{\begin{eqnarray}}
\newcommand{\eea}{\end{eqnarray}}
\newcommand{\beq}{\begin{equation}}
\newcommand{\eeq}{\end{equation}}
\newcommand{\nnu}{\nonumber}
\newcommand{\cH}{{\cal H}}
\newcommand{\cM}{{\cal M}}
\newcommand{\cO}{{\cal O}}

\newcommand{\sbar}{{\overline s}}

\newcommand{\nubar}{{\overline \nu}}

\newcommand{\ba}{\begin{array}}
\newcommand{\ea}{\end{array}}
\newcommand{\lsim}{\stackrel{<}{_\sim}}
\renewcommand{\Re}{{\rm Re}}
\renewcommand{\Im}{{\rm Im}}

\def\npps#1#2#3{   {\it Nucl. Phys. Proc. Suppl. }{\bf #1} (#2) #3}

\def\plb#1#2#3{    {\it Phys. Lett. }{\bf B #1} (#2) #3}

\def\prd#1#2#3{    {\it Phys. Rev. }{\bf D #1} (#2) #3}

\def\prl#1#2#3{    {\it Phys. Rev. Lett. }{\bf #1} (#2) #3}

\def\ijmpa#1#2#3{  {\it Int. J. Mod. Phys. }{\bf A #1} (#2) #3}
\def\sjnp#1#2#3{   {\it Sov. J. Nucl. Phys. }{\bf #1} (#2) #3}

\begin{document}

\thispagestyle{empty}
\begin{flushright}
CERN--TH/2002--053 \\
hep-ph/0203044 
\end{flushright}
\vskip 3.0 true cm 

\begin{center}
{\Large\bf T-odd correlations in charged $K_{\ell 4}$ decays} 
 \\ [25 pt]
{\sc {\sc Alessandra Retico} 
 \\ [25 pt]
{\sl  Theory Division, CERN, CH-1211 Geneva 23, Switzerland} \\ [5 pt]
{\sl  INFN, Sezione di Roma and Dipartimento di Fisica, \\
           Universit\`a di Roma ``La Sapienza'', P.le A. Moro 5, I-00185 Rome, Italy} \\ 
[25 pt]  }
   {\bf Abstract} \\
\end{center}\noindent
We analyse the sensitivity to physics beyond the SM 
of T-odd correlations in $K_{\ell 4}$ decays, which do not 
involve the lepton polarization. We show that a combined analysis 
of $K^+_{\mu 4}$ and  $K^-_{\mu 4}$ decays can lead to new 
constraints about CP violation in $\Delta S=1$ charged-current 
interactions, complementary to those 
obtained from the transverse muon polarization in $K_{\mu 3}$
and of comparable accuracy. 

\bigskip
\def\thefootnote{\arabic{footnote}}
\setcounter{footnote}{0}

\paragraph{I.}
Charged kaon decays provide an ideal framework to study 
{\em direct} CP violation, or CP violation of pure   
$\Delta S=1$ origin \cite{DI}. Within the Standard Model (SM), this 
is expected to be very suppressed (at the $10^{-5}$--$10^{-6}$ level) 
in non-leptonic decays, such as $K^\pm \to (3\pi)^\pm$, and absolutely negligible
in charged-current semileptonic decays, such as $K^\pm_{\ell 3}$ 
and $K^\pm_{\ell 4}$. These processes therefore 
offer a clean testing ground for non-standard scenarios,
which could soon be studied with high precision 
both at KLOE \cite{KLOE} and NA48 \cite{NA48}, 
where a large statistics of both $K^+$ and $K^-$ 
fluxes is expected. 

The simplest CP-violating observable in charged kaon decays 
is the charge asymmetry of the decay widths, 
$\Gamma(K^+ \to f) - \Gamma(K^- \to \bar f)$,
or, equivalently, the charge asymmetry of any T-even differential 
decay distribution. However, these observables can be 
different from zero only in the presence of at least two weak 
amplitudes with different final-state interaction (FSI) phases. 
In order to avoid this condition, which provides a severe 
model-independent suppression of charge asymmetries  
in non-leptonic modes \cite{DI} (and would prevent the  
study of CP violation in $K_{\ell 3}$), CP violation 
can be tested by means of T-odd correlations. 
The most well-known example of such correlations
is the transverse muon polarization in $K^+_{\mu 3}$, 
$P^T_\mu = \langle \vec{\sigma}_\mu \cdot (\vec{p}_\mu \times \vec{p}_\pi)/
\vert  \vec{p}_\mu \times \vec{p}_\pi \vert \rangle$,
which has been extensively discussed in the literature 
\cite{Kl3th} (see also \cite{Diwan} and references therein)
and has been found to be consistent with zero 
below the $10^{-2}$ level \cite{Kl3exp}. 

Contrary to the $K_{\ell 3}$ case, $K_{\ell 4}$ modes offer 
the possibility to study T-odd correlations even without
measuring the lepton polarization, which would not be 
accessible in experiments such as KLOE and NA48.
In particular, the expectation value of the triple product 
$\vec{p}_\ell \cdot (\vec{p}_{\pi_1} \times \vec{p}_{\pi_2})$
provides a T-odd correlation accessible in $K_{\ell 4}$ decays only.
In general a non-vanishing T-odd correlation does not 
necessarily imply an evidence of CP violation: this is true only 
in the absence of FSI phases.  The latter are known to be extremely 
suppressed in $K^+_{\mu 3}$ \cite{Zit}, but $\cO( 10^{-1})$ 
strong phases are certainly present 
in $K_{\ell 4}$. However, this problem can be overcome 
by the analysis of both $K^+$ and $K^-$ modes, whose 
T-odd correlations can be combined to construct a pure 
CP-violating observable. Therefore this study
appears particularly appealing  for high statistics 
experiments with both $K^+$ and $K^-$ fluxes, such as 
KLOE and  NA48. 

In this letter we present a general analysis of the 
above-mentioned T-odd correlation in $K_{\ell 4}$ decays. We shall 
compare the sensitivity of this observable with respect to 
$P^T_\mu$ in testing the most general structure of $\Delta S =1$
charged currents. As we shall show, $K_{\ell 4}$ decays 
can be used to extract new information about 
non-standard sources of CP violation. 

\paragraph{II.} Assuming purely left-handed neutrinos, 
the most general dimension-six effective Hamiltonian 
relevant to $K_{\ell 3}$ and $K_{\ell 4}$ decays 
can be written as 
\beq
\cH^{\rm eff}  =2\sqrt{2} G_F V^*_{us} 
\sum_{i}C_{i} O_{i} ~+~{\rm h.c.}
\label{eq:H_eff}
\eeq 
where the operator basis is given by 
\beq
\ba{rclrcl}
O^V_L &=&  \sbar_L \gamma^\mu u_L \nubar_L\gamma_\mu \ell_L~, & \qquad 
O^V_R &=&  \sbar_R \gamma^\mu u_R \nubar_L \gamma_\mu \ell_L~, \\ 
O^S_L &=&  \sbar_R u_L \nubar_L \ell_R~, & \qquad \qquad
O^S_R &=&  \sbar_L u_R \nubar_L \ell_R~, \nnu \\ 
O^T_L &=&  \sbar_R \sigma^{\mu\nu} u_L \nubar_L \sigma_{\mu\nu}\ell_R~, &
O^T_R &=&  \sbar_L\sigma^{\mu\nu} u_R \nubar_L \sigma_{\mu\nu} \ell_R~.  \ea
\label{eq:sc_ops}
\eeq
Within the SM, all the $C_i$ are negligible except for $C^V_L=1$ and, 
in general,
all the $C_i$ can be assumed to be real in the absence of CP violation.

The hadronic matrix elements relevant to the $K^+_{\mu 3}$
decay can be written as
\bea
\langle  \pi^0(p_\pi)| \sbar \gamma^\mu  u |K^+(p_K)\rangle 
 &=&\frac{1}{\sqrt{2}}
        [(p_k+p_\pi)^\mu f_+(q^2)
        +(p_k-p_\pi)^\mu f_-(q^2)   ]~, \nnu \\
\langle\pi^0(p_\pi)| \sbar u |K^+(p_K)\rangle &=&
\frac{1}{\sqrt{2}} m_K B_S (q^2)~, \nnu \\
\langle\pi^0(p_\pi)| \sbar\sigma^{\mu\nu} u |K^+(p_K)\rangle &=&
    i \frac{B_T(q^2)}{\sqrt{2} m_K} 
(p_K^\mu p_\pi^\nu -p_K^\nu p_\pi^\mu)~,
\eea
where $q^2=(p_k-p_\pi)^2=(p_\ell+p_\nu)^2$. The form 
factors are all real (neglecting tiny electromagnetic 
effects \cite{Zit}) and their dependence on 
$q^2$ starts only beyond lowest order in
the chiral expansion (see e.g. Ref.~\cite{bijnens}). 
Concerning scalar and vector form factors, the lowest 
order expressions, which already provide a reasonable 
approximation for our purposes, are given by\footnote{~For 
the scalar form factor see e.g. Ref.~\cite{Colangelo} and for
a complete next-to-leading [${\cal O}(p^4)$] evaluation of the vector 
and axial form factors
see e.g. Ref.~\cite{bijnens}.} 
\beq
f_+ = 1~, \qquad f_- = 0~, \qquad B_S = \frac{m_{K}}{m_s + m_u} \approx 4~.
\eeq
The tensor form factor, which cannot be directly related to 
observable quantities or fundamental parameters (such as $m_s$), 
has recently been measured on the lattice obtaining 
$B_T= 1.23 \pm 0.09$ (with a negligible $q^2$ dependence) \cite{SPQR}, 
in good agreement with the estimate $B_T \approx 1$ made in \cite{CIP}.
As the scalar and the tensor operators have non-null anomalous dimensions,
$B_S$ and $B_T$ scale according to the renormalization 
group equations. The values given above 
refer to a renormalization scale around 2 GeV.
With these notations the transverse muon polarization 
in $K^+_{\mu 3}$ is given by 
\beq
P^T_\mu =  \frac{1}{2} \frac{m_\mu}{m_K} \left<  \,
\frac{ {\vec p}_\nu\times {\vec p}_\mu}{E_\nu E_\mu} \,  {\rm Im}\,\xi  \,
\right>\, ,
\label{eq:xidef}
\eeq
where
\beq
\xi =  \left[ \left( 1 - \frac{f_-}{f_+} \right) +  
\frac{m_K }{ m_\mu} \frac{C^S_+ B_S}{C^V_+f_+} + 
\frac{2 p_K (p_\nu - p_\mu)}{ m_\mu m_K }\frac{C^T_+ B_T}{C^V_+f_+} \right] 
\times \left[ 1  -  \frac{m_\mu}{m_K} \frac{ C^T_+ B_T}{C^V_+f_+}  \right]^{-1}
\label{eq:xith}
\eeq
and $C^i_\pm =C^i_R \pm C^i_L$. 
Neglecting the kinematical dependence of ${\rm Im}\, \xi$, the 
experimental information on $P^T_\mu$ 
implies ${\rm Im}\, \xi = - 0.013 \pm 0.016$ or  $|{\rm Im}\, \xi| 
< 0.033$~(90\% C.L.) 
\cite{Kl3exp}. 
Because of the chiral enhancement of the scalar matrix element, the most 
stringent constraints are obtained on the CP-violating phases of 
$C^S_{L,R}$: $|{\rm Im}\, C^S_{L,R}| \lsim 10^{-3}$ (assuming $C^V_{+}$
to be real). This holds in the absence of cancellations between 
left-handed and right-handed terms. However, since the 
parity-violating combination ${\rm Im}\, C^S_-$ is 
not accessible in $K_{\mu 3}$, in principle larger values 
of ${\rm Im}\, C^S_{L}$ and ${\rm Im}\, C^S_{R}$, such that  
$|{\rm Im}\, C^S_-| \gg |{\rm Im}\, C^S_+|$, cannot be excluded yet.

\paragraph{III.} The hadronic matrix elements 
appearing in $K_{\ell 4}$ decays, within the SM, 
can be parametrized as \cite{bijnens}
\bea
\langle\pi^+(p_+) \pi^-(p_-)| \sbar \gamma^\mu u |K^+(p_K)\rangle &=&
-\frac{H}{m^3_K}\epsilon_{\mu\nu\rho\sigma}L^\nu P^\rho Q^\sigma\nnu~, \\
\langle\pi^+(p_+) \pi^-(p_-)| \sbar \gamma^\mu \gamma_5 u |K^+(p_K)\rangle &=&
\frac{-i}{m_K}[P_\mu F + Q_\mu G + L_\mu R]~,
\eea
where $P_\mu=(p_+ + p_-)_\mu$, $Q_\mu=(p_+ - p_-)_\mu$, 
$L_\mu=(p_\ell + p_\nu)_\mu$ and 
the $\cO(1)$ form factors $H$, $F$, $G$, $R$ are complex functions of 
three independent kinematical variables. 
A precise experimental determination of these form
factors has recently been obtained by the BNL--E865
Collaboration \cite{E865}, updating the old result by 
Rosselet {\em et al.} \cite{Rosselet}. 

Concerning the matrix element of non-standard contributions, 
not determined by experiments, we shall restrict our attention to 
those not vanishing at lowest order in the chiral expansion (expected 
to be dominant):
\bea
\langle\pi^+(p_+) \pi^-(p_-)| \sbar\gamma_5 u |K^+(p_K)\rangle &=&  i S  \nnu~, \\
\langle\pi^+(p_+) \pi^-(p_-)| \sbar\sigma^{\mu\nu}\gamma_5
 u |K^+(p_K)\rangle &=&
\frac{T}{m^2_K} [p^\mu_{-} (p^\nu_{+} + p^\nu_{K})
-p^\nu_{-} (p^\mu_{+} + p^\mu_{K})]~.
\eea
The lowest-order results for $S$ (see e.g. Ref.~\cite{bijnens}) 
and $T$ (see Ref.~\cite{CIP}) are given by 
\beq
S = \frac{\sqrt{2} m_K }{3 F_\pi  } B_S~, \qquad 
T = \frac{m_K }{ \sqrt{2} F_\pi } B_T~,
\eeq
but of course also these form factors 
acquire a non-trivial phase  beyond lowest order.

The matrix element is then given by
\bea
\cM &=& \sqrt{2} G_F V^*_{us} 
\left\{ \left[
- C^V_+ \,H \epsilon^{\mu\nu\rho\sigma}\frac{L_\nu P_\rho Q_\sigma}{m^3_K}
- i  C^V_-  \frac{P^\mu}{m_K} \left(F+ 3 \frac{ m_l}{ m_K } 
\frac{C^T_-}{C^V_-} \, T  \right)  \right.\right. \nnu \\
&-& \left. i  C^V_- \frac{Q^\mu}{m_K} \left( G + \frac{ m_l}{m_K}  
 \frac{C^T_-}{C^V_-}\, T  \right) 
- i  C^V_- \frac{L^\mu}{m_K} \left( R+ 2 \frac{ m_l}{ m_K } 
 \frac{C^T_-}{C^V_-} \,T \right) \right]
\,{\bar u}_L (p_\nu) \gamma_\mu v_L (p_l) \nnu\\
&+& i  \left[  C^S_-  \, S    
- 2 C^T_-  \,T \frac{  (p_l-p_\nu)(p_+ + p_K)}{m^2_K} \right]\,
{\bar u}_L (p_\nu) v_R (p_l)  \nnu\\
&-& \left. 4  C^T_-  \,T   \, \frac{p^\mu_+ p^\nu_K}{m^2_K}  
\,{\bar u}_L (p_\nu) \sigma_{\mu\nu} v_R (p_l) 
 \right\}~.
\eea
The interference between vector and scalar 
terms leads to the following T-odd term in the 
matrix element squared, summed over lepton polarizations:
\bea
 \sum_{\rm pol}\cM^2 (K^+)  \Big|_{\rm T-odd} &=& 
-  16 G^2_F |V_{us}|^2 \frac{m_l}{ m_K^2 } 
\Im\left\{ C^V_+ H  ( C^S_-S)^* \right.
\nnu\\  && \left. \times  \left[ 1- 2 \frac{(C^T_- T)^*}{(C^S_- S)^*}
\frac{  (p_l-p_\nu)(p_+ + p_K)}{m^2_K} \right] \right\}~ 
 ({\vec p}_+\times {\vec p}_-) \cdot{\vec p}_{l^+} ~.
\label{eq:Toddp}
\eea
As can be noted, if the form factors $S$, $T$  and  $H$ were all 
relatively real, this term would vanish in the absence of CP violation. 
In the CP-conjugate process,  $K^-\rightarrow\pi^+\pi^- \ell^- \nu$,
the analogous interference term is given by 
\bea
 \sum_{\rm pol}\cM^2 (K^-)  \Big|_{\rm T-odd} &=& 
+  16 G^2_F |V_{us}|^2 \frac{m_l}{ m_K^2 } 
\Im\left\{ (C^V_+)^* H C^S_-(S)^*\right.
\nnu\\  && \left. \times \left[ 1 - 2 \frac{C^T_- (T)^*}{ C^S_- (S)^*}
\frac{(p_l-p_\nu)(p_- + p_K)}{m^2_K} \right] \right\}~ 
 ({\vec p}_+ \times {\vec p}_-) \cdot{\vec p}_{l^-} ~.
\label{eq:Toddm}
\eea
Thus if we neglect the highly suppressed tensor term quadratic in 
$({\vec p}_+ - {\vec p}_-)$, the sum of these two T-odd 
correlations becomes a clear CP-violating observable. 

The unpolarized differential decay rate 
of $K_{\ell 4}$ decays can generally be written as 
\bea
d\Gamma = G^2_F|V_{us}|^2 N(s_\pi,s_l)\, 
J_5(s_\pi,s_l,\theta_\pi,\theta_l,\phi)
ds_\pi ds_l d(\cos\theta_\pi) d(\cos\theta_l) d\phi~,
\eea
where the five independent kinematical variables are 
defined as in \cite{bijnens} and all the dynamical 
information is encoded in $J_5$. Furthermore,
from the experimental point of view it is convenient 
to decompose $J_5$ as \cite{PaisT}
\bea
J_5&=& 2 (1-z_l) [I_1 + I_2 \cos 2 \theta_l + I_3 \sin^2 \theta_l \cos 2\phi
+ I_4  \sin 2 \theta_l \cos \phi + I_5  \sin \theta_l \cos \phi+\nnu\\
&+& I_6  \cos \theta_l + I_7\sin \theta_l \sin\phi 
+ I_8 \sin 2 \theta_l \sin\phi + I_9 \sin^2 \theta_l \sin 2\phi]~,
\eea
showing the explicit dependence on the leptonic variables $\theta_l$ 
and $\phi$. The T-odd correlations (\ref{eq:Toddp}) and (\ref{eq:Toddm}) 
contribute to $I_7$ and, neglecting the suppressed contribution
proportional to $C_-^T$, the CP-violating combination 
$[I_7(K^+_{\ell 4}) + I_7(K^-_{\ell 4})]$ is given by 
\bea
I_7(K^+_{\ell 4}) + I_7(K^-_{\ell 4})
&=& \frac{m_\ell}{m_K } \frac{\lambda^\frac{1}{2}(m^2_K,s_\pi,s_\ell)}{(1-z_l)}
\gamma \sqrt{s_\ell}\sqrt{s_\pi-4 m_\pi^2} 
\sin \theta_\pi \nnu\\
&&\times ~\Re (H S^*) ~|C^V_+|^2~ \Im \left(\frac{C^S_-}{C^V_+} \right)
\label{eq:final} 
\eea
$[\gamma=
(1-\lambda(m^2_K,s_\pi,s_\ell)/(m^2_K-s_\ell+s_\pi)^2)^{-\frac{1}{2}}]$.
As expected, this observable is suppressed by the lepton mass and 
in practice can be studied only in the muon case. In these channels, 
$I_7$ would reach 20\% of the dominant form factor ($I_1$) for 
$\Im ( C^S_-/ C^V_+ )\sim 1 $. Thus in high-statistics experiments 
the possible bounds on $|\Im\, C^S_-|$ could become competitive with 
the $\cO(10^{-3})$ bounds on $|\Im\, C^S_-|$, derived from 
$P^T_\mu(K^+_{\mu3})$.
Moreover, we recall that the two bounds are in principle independent, 
since we cannot exclude a priori a scenario where 
$|{\rm Im}\, C^S_-| \gg |{\rm Im}\, C^S_+|$.

\paragraph{IV.}
In this letter we have presented a general analysis of 
T-odd correlations accessible in $K_{\ell 4}$ decays,
without the measurement of lepton polarization. 
As we have shown, in general these are not clean
CP-violating observables, due to the large FSI phases affecting
$K_{\ell 4}$ amplitudes. However, combining information 
from $K^+$ and  $K^-$ modes it is possible to construct a 
clean CP-violating observable, sensitive to non-standard 
$\Delta S =1$ charged-current interactions. 
Future high-statistics experiments on  $K_{\ell 4}$ decays
could use this observable to put new constraints on 
exotic non-standard scenarios, such as R-parity 
violating supersymmetry, complementary to those 
obtained from $K_{\ell 3}$ decays \cite{Kl3th}.

\section*{Acknowledgements}

I am very  grateful to G. Isidori for interesting suggestions and discussions.

\end{document}